\documentclass[]{aa}
\usepackage{graphicx}
\usepackage{times}
\usepackage{natbib}
\bibpunct{(}{)}{;}{a}{}{,} 

\newcommand{\msun}{${M}_\odot$}

\begin{document}
  \title{The post-AGB evolution of AGB mass loss variations}

  \author{Rowin Meijerink\inst{1} \and Garrelt Mellema\inst{1} \and Yvonne Simis\inst{2,1}}

  \institute{Sterrewacht Leiden, P.O. Box 9513, 2300 RA, Leiden, 
    The Netherlands\\
    \email{meijerin@strw.leidenuniv.nl, mellema@strw.leidenuniv.nl,\\simis@strw.leidenuniv.nl}
    \and
    Astrophysikalisches Institut Potsdam, An der Sternwarte 16, 
    D-14482 Potsdam, Germany }

  \offprints{R.~Meijerink}

  \date{Received 17 December 2002 / Accepted 30 April 2003}

  \abstract{We present new numerical hydrodynamical modelling of the evolution
    of Asymptotic Giant Branch (AGB) mass loss fluctuations during the
    post-AGB/Planetary Nebula phase. These models show that after ionization,
    the observable effects of the mass loss fluctuations disappear in a few
    thousand years, consistent with the fact that only few PNe have been found
    to be surrounded by `rings'. We derive the observational characteristics
    of these rings, and compare them to reported observations of the rings
    around NGC 6543, finding a good match of emission properties and line
    shapes. We predict small variations in the observable electron
    temperatures.

 }

  \maketitle
%

\section{Introduction}

High resolution imaging with the Hubble Space Telescope {\it HST}\/ have
revealed the presence of so-called `rings' around Planetary Nebulae (PNe) and
proto-PNe. Examples can be found around the proto-PNe CRL 2688 \citep[Egg
Nebula]{sahaietal98}, IRAS~17441-2411 \citep{Suetal98}, IRAS~17150-3224
\citep{Kwoketal98}, IRAS 16594-4656, IRAS~20028+3910 \citep{Hrivnaketal01},
and the PNe NGC 7027, (where they show up in scattered light), NGC~6543 and
Hb~5 \citep{TerzianHajian00}. Inspection of {\it HST}\/ archival images of
NGC~3918 also reveal ring-like structures in its halo, confirmed by
ground-based data \citep{Corradi}. Similar rings are observed around
one AGB star, the extreme Carbon star IRC+10216 \citep{MauronHuggins}.

The name `rings' is somewhat of a misnomer since these are three-dimensional
structures. However, the term is now commonly used and we will adher to this
convention. Although never completely closed, the rings show an overall
spherical appearance, and seem to be associated with a time when the mass
loss from the AGB star was roughly spherical.  Interestingly, they tend to
sit right around the aspherical nebular structures, suggesting they are
connected to the phase just before the onset of asphericity in the system.

Their origin has been attributed to three different processes.
\citet{MastroMorris99} showed that binary interaction can produce ring-like
structures, both in and perpendicular to the orbital plane. In this case the
rings are actually the spiral waves produced by the presence of the companion
star. \citet{Guille2001} reproduce the ring features by invoking a magnetic
cycle, similar to that in the Sun, for the AGB star. The period is then set
by the period of this magnetic activity cycle, which is a parameter in their
model. Frank (private comm) has pointed out that any magnetic dynamo cycle in
AGB stars would have a much shorter period than the one corresponding to the
rings. The third model comes out of a study of the dust-driven winds of AGB
stars. \citet{Simisetal2001} showed that intricate coupling between dust and
gas, leads to cycles in which predominantly either large or small grains are
produced, which then produce periodic variations in the wind density and
velocity. The period depends on the sound travel time of the subsonic wind
region, and is found to lie close to the observed values for the rings.

In case the rings are ionized it is possible to use plasma diagnostics to
derive the temperature and density in the rings. To date this has not been
done in detail. \citet{Balicketal2001} estimated some properties of the rings
around NGC~6543 on the basis of {\it HST}\/ images and limited spectroscopic
data.  \citet{Hyungetal2001} used {\it HST\/} images taken with {\it WFPC2\/}
in narrow band filters near the lines of [O~III] 5007\AA\ and 4363\AA\ to
find the electron temperature in the rings of the same object. Although
\citet{Balicketal2001} found no indication for unusual electron temperatures
in the rings, \citet{Hyungetal2001} claim to find substantially higher
electron temperatures, indicating the presence of shocks. Both methods have
their limitations, and the result remains
inconclusive. \citet{Balicketal2001} derived density variations from
H$\alpha$ emission variations, and found them to be consistent with
individual shells with negligible amounts of material between them.

Both \citet{Bryceetal92} and \citet{Balicketal2001} report that emission
lines in the rings area of NGC~6543 are unusually broad, with FWHM of around
30~km~s$^{-1}$, for which they could offer no explanation.
\citet{Hyungetal2001} presented a simple model for explaining the putative
higher electron temperatures, as well as the broadened emission
lines. Following the evolution of sinusoidal density and velocity variations
using a numerical hydrodynamic model including the effects of ionization,
they could reproduce temperatures and broadened lines. However, the model
only included the rings, not the core nebula, nor did it follow the full
evolution of the rings, and in order to keep the velocity field under
control, unreasonably high velocities had to be used.

Studying the evolution of the rings from the AGB to the PN phase is
interesting for several reasons. Firstly, one may be able to find differences
between the three proposed models. Secondly, it can help us understand
whether the limited number of PNe found to have rings is intrinsic, e.g. 
because ionization changes the properties of the rings, or rather
due to lack of high resolution observations. Also, it will provide insight in
the connection between the AGB mass loss variations and the ionized PN rings.

In this paper we study the evolution of mass loss variations on the AGB
through the proto-PN phase to the PN phase. For this we will use an improved
version of the model of \citet{Simisetal2001} to produce the AGB mass loss
variations, and an improved version of the model used in \citet{RHPNIII} to
follow their evolution during the post-AGB phase. Although we pick one model
for the origin of the mass loss variations, we will show that most of the
results have relevance for the other models too.

The layout of the paper is as follows. In Sect.~2 we describe the numerical
methods used to produce the mass loss variations and to study their fate from
the AGB through the post-AGB phase. Section~3 presents the results of a
typical simulation, for which we derive observables in Sect.~4. In Sect.~5 we
discuss the results and how they increase our understanding of the origin and
evolution of the rings. Our conclusions are summed up in Sect.~6.

\section{Two models}

In order to model the evolution of AGB mass loss variations we first need to 
produce them. To keep the model as physical as possible, we produce the mass
loss fluctuations using a slightly improved version of the two-fluid dust-gas
code described by \citet{Simisetal2001}. These are then used as input for a
gas-only hydrodynamics code which can treat the detailed ionization,
heating and cooling processes in step with the evolution of the star. We will
now describe these codes in some more detail.

\subsection{A dust-driven AGB wind model}

We provide here a short description of the two-fluid hydrodynamics code which
we use to model the AGB. For a more detailed description of the code and the
implementation of the two-fluid flow in particular, the reader is referred to
\citet {Simisetal2001,Simisetal2003}.
The new aspect of this code is the fact that it combines time-dependent
hydrodynamics with two-fluid flow and a self consistent description of the
dust (including nucleation and growth). It turns out that the interaction of
these physical mechanisms is critical for the formation of the shells, which
are a result of alternating low- and high-mass loss episodes of outflow.  The
mass loss rate is low when the momentum transfer from grains to gas is
inefficient. This is the case when the gas has a relatively low velocity
while the dust grains move relatively fast (i.e. they have a high drift
velocity). This results in a rapid passage of the grains through the zone in
which grain growth is efficient, and hence in somewhat smaller grains. As a
consequence, the drag force exerted on the gas is relatively weak. This leads
to a decrease of the gas density and hence an even larger drift velocity,
because the frequency of gas-grain collisions decreases. So, during the low
mass-loss phase the average grain size gradually decreases whereas its drift
speed increases.  In a phase of high mass loss, the rate of momentum transfer
from the radiatively accelerated grains to the gas is high. This results in a
large velocity of the gas and a low drift velocity. Because the grains move
relatively slow through the gas they have sufficient time to grow. The
average grain size during this phase is relatively large, which causes an
efficient momentum transfer. This enables a high gas density, a gradual
growth of the grains and a further decrease of their drift speed.  The
transition between the two phases is triggered by the sudden increase of the
drag force at the end of the low mass loss phase, which is a result of the
quadratic appearance of the (increasing) drift velocity in the drag
force. The sudden increase of the drag force leads to an increasing momentum
transfer per collision and it prevents the ongoing decrease of the collision
rate that was a result of the gradually decreasing grain size. This starts
the high mass loss phase. Curiously, it also triggers the onset of the next
low mass loss phase: due to the suddenness of the transition, an outward
shock develops. At the same time, a rarefaction wave moves towards the
stellar surface. This wave will dilute the gas, thereby again creating a
condition for inefficient grain growth, small grains, low collisional cross
sections and hence a low mass loss rate. Hence, the characteristic time scale
for the mass loss fluctuations is the time used by the rarefaction wave to
cross the dust forming part of the subsonic region of the outflow.

The code is a time dependent, Eulerian, explicit hydrodynamics code, based on
the FCT/LCD algorithm assuming spherical symmetry
\citep{Boris:1976,Icke:1991}.  Physical processes taken into account in the
code involve (equilibrium) gas chemistry, nucleation and growth of grains
\citep{Gailetal84, GailSedlm88, DorfiHofner91}, semi-analytical radiative
transfer \citep{Fleischeretal92} and the behaviour of grains in collision
with gas particles. The temperature is calculated assuming local radiative
equilibrium, following \citet{Lucy76}. Grain-gas collision is implemented to
provide a realistic description of the momentum transfer between the two
fluids. This enables a full two-fluid description, i.e. no assumptions about
the motion of the grains relative to the gas need to be made. The extra
degree of freedom introduced thereby turns out to be essential for the
formation of quasi periodic shells: when comparing two otherwise identical
model calculations, the (two-fluid) model in which grains can drift with
respect to the gas does produce the shells, whereas the single fluid model,
i.e. the model in which grains are always assumed to have the same velocity
as the gas, has a more or less stationary outflow.

\begin{figure}
\centerline{\includegraphics[angle=90,height=67mm,clip=]{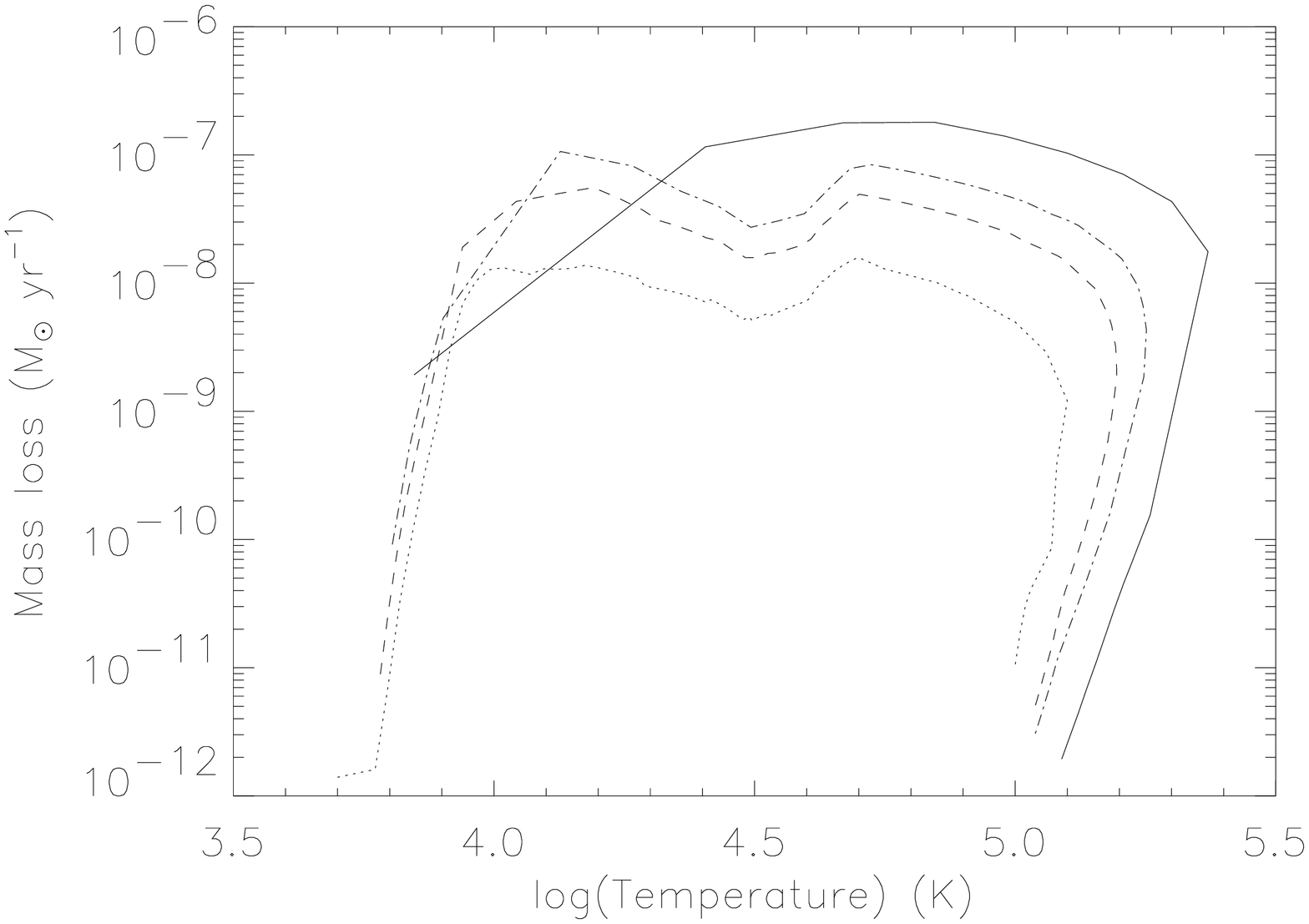}}
\centerline{\includegraphics[angle=90,height=67mm,clip=]{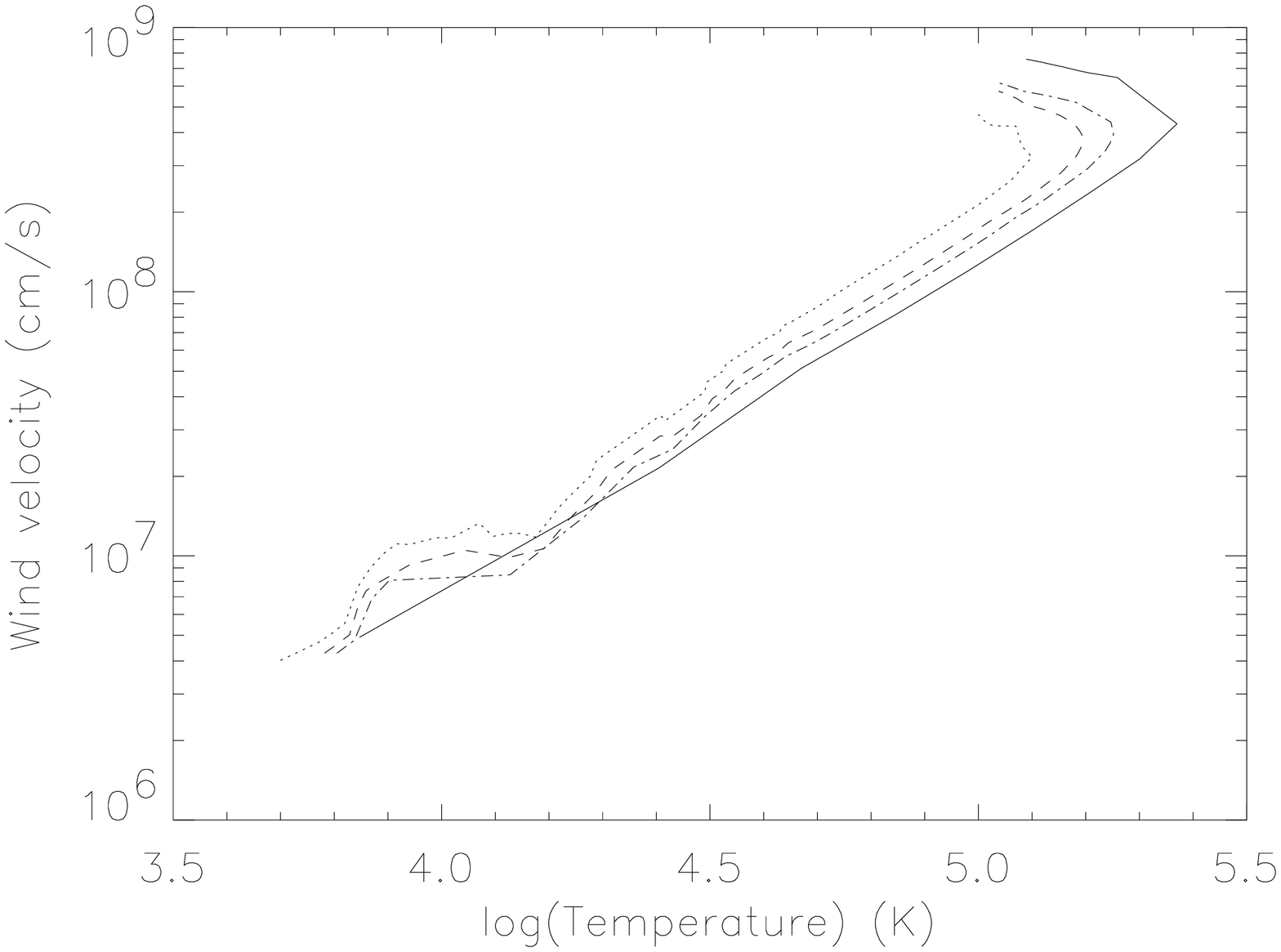}}
\caption{The mass loss of the star and the velocity of the wind as a function 
of effective temperature, for post-AGB masses of 0.565 M$_\odot$ (dotted), 
0.605 M$_\odot$ (dashed), 0.625 M$_\odot$ (dashed dotted) and 0.696 M$_\odot$ 
(solid).}
\label{masslossvel}
\end{figure}

\subsection{A post-AGB model}

The numerical method we use to follow the subsequent evolution in the
post-AGB phase, descends from the method used by \citet{RHPNIII}. This
code
solves the hydrodynamic equations in one-dimension, using the spherical
radial coordinate. The solver of choice is the `Roe solver'
\citep{Roe81,EulMel95}, a second-order accurate, approximate Riemann
solver.

The atomic physics is treated according to the approach described in
\citet{MellemaLundqvist2002} ({\it DORIC}), but with the addition of
photo-ionization from an evolving source. This means that we follow the
time-dependent ionization of all ions of H, He, C, N, O, and Ne, taking into
account collisional ionization, photo-ionization, radiative and dielectronic
recombination and charge exchange with hydrogen. The local density and
temperature of the ions is used to calculate the local non-equilibrium
cooling rate. The heating rate is given by photo-ionization heating of H and
He.

The transfer of ionizing radiation is treated according to the method
described in \citet{RH-PN1} and \citet{Photoclump}, i.e.~using the three
frequencies corresponding to the ionization thresholds of H$^0$, He$^0$, and
He$^+$, but taking the frequency dependence of the cross sections into
account when calculating the integrals over the spectral energy
distribution. At all times, the star is assumed to radiate as a black body
with the effective temperature given by the stellar evolution model used (see
below).

The outer boundary is an outflow condition, the inner boundary condition
is provided by the evolving stellar wind in the post-AGB phase.

\subsubsection{The post-AGB wind}

During the post-AGB phase a star can reach temperatures over 10${^5}$~K.
The bulk of the radiation is emitted in the UV, where the
outer layers have an abundance of absorption lines. \citet{CAK1975} found
that for hot stars, absorption lines dominate in transferring momentum from
the radiation to the gas, and the consequent stellar wind is therefore called
`line-driven'. The theory of line-driven winds is well developed and tested.
Here we use the analytical approximation derived by
\citet{Kudritzkietal1989}, which for a given effective temperature and
luminosity gives the stellar wind mass loss rate and terminal velocity.

The mass loss rate of a line-driven wind is proportional to the luminosity
and the number of absorbing lines. The acceleration of the atoms is
proportional to a so called force multiplier, which is approximated by a
simple function:
\begin{equation}
M(t)=k\ t^{-\alpha}(10^{-11}n_e/W)^\delta,
\end{equation} 
where $t$ is a dimensionless optical depth parameter, $k$, $\alpha$ and
$\delta$ force multiplier parameters and $W$ the dilution factor. The
acceleration also depends on the velocity field in the wind as a function of
radius. This velocity field is given by:
\begin{equation}
v(r)=v_\infty(1-\frac{R_*}{r})^{-\beta}
\end{equation}
Apart from the stellar luminosity and effective temperature, the model also
needs values for these force multiplier parameters, $k$, $\alpha$, $\beta$
and $\delta$, as well as for the stellar mass, the helium abundance, and the
average number of electrons produced by helium.

The luminosity, mass and effective temperature are taken from
\citet{Bloecker1995}, who presented a compilation of evolutionary tracks for
post-AGB masses in the range 0.53 to 0.94 M$_\odot$. Three of these tracks
were originally calculated by \citet{Schoenberner1983}.
The force multiplier parameters we take
from the compilation in \citet[Table 8.2]{Lamersetal1999}.

With these parameters the wind depends strongly on the stellar effective
temperature.  When the star just enters the post-AGB phase, its effective
temperature is low, and the mass loss rate and wind velocity are low too. As
the effective temperature goes up, the mass loss rate increases, levelling
off to a roughly constant rate from about 10000~K.  At this point the number
of effective lines reaches a maximum. When nuclear burning stops, the
luminosity of the star starts to drop and the mass loss rate goes down, but
the velocity of the wind keeps rising.  Figure~\ref{masslossvel} shows the
evolution of the wind as a function of stellar effective temperature for four
different evolutionary tracks.

\begin{figure}
\includegraphics[angle=90,height=67mm,clip=]{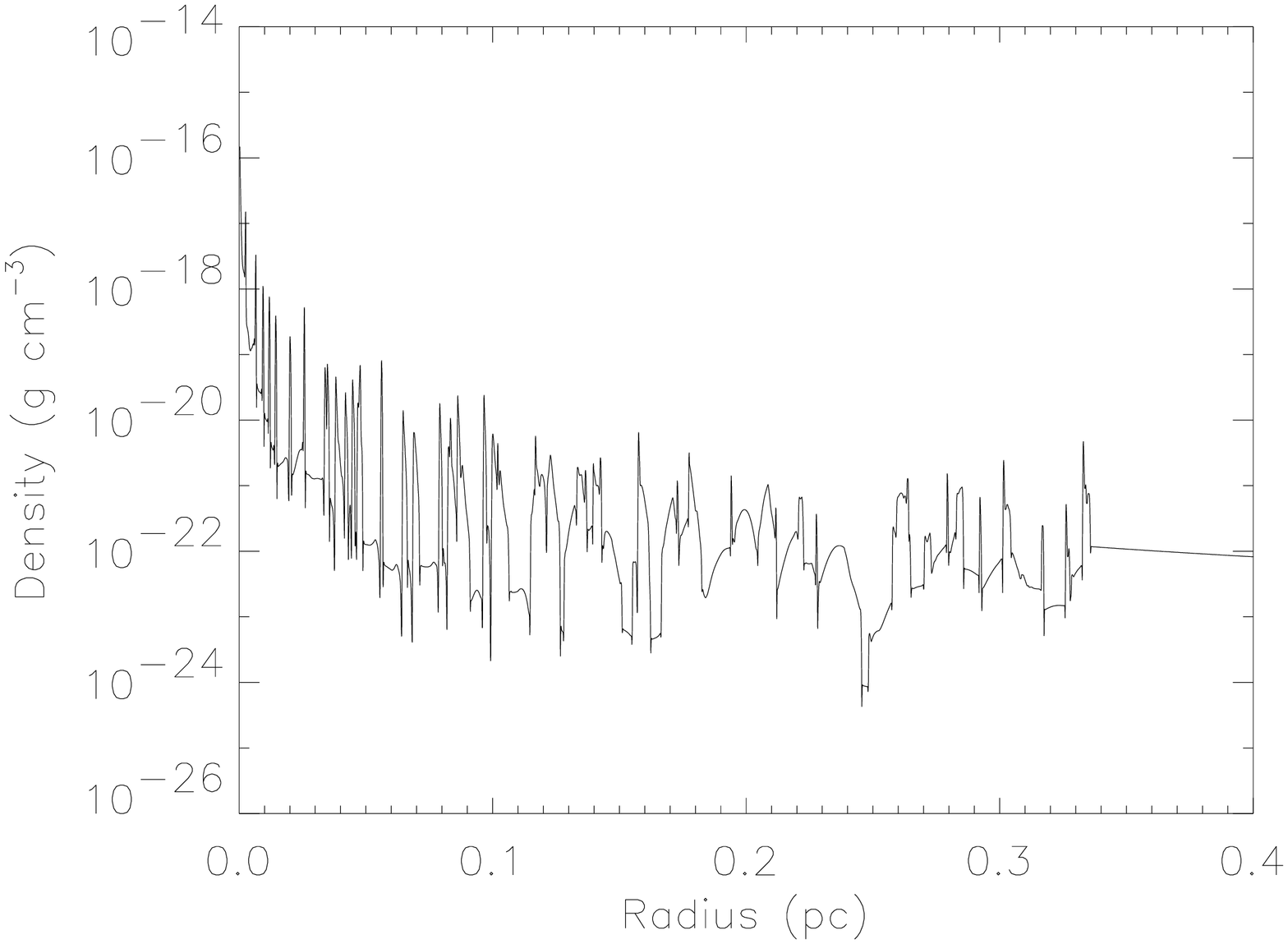}
\includegraphics[angle=90,height=67mm,clip=]{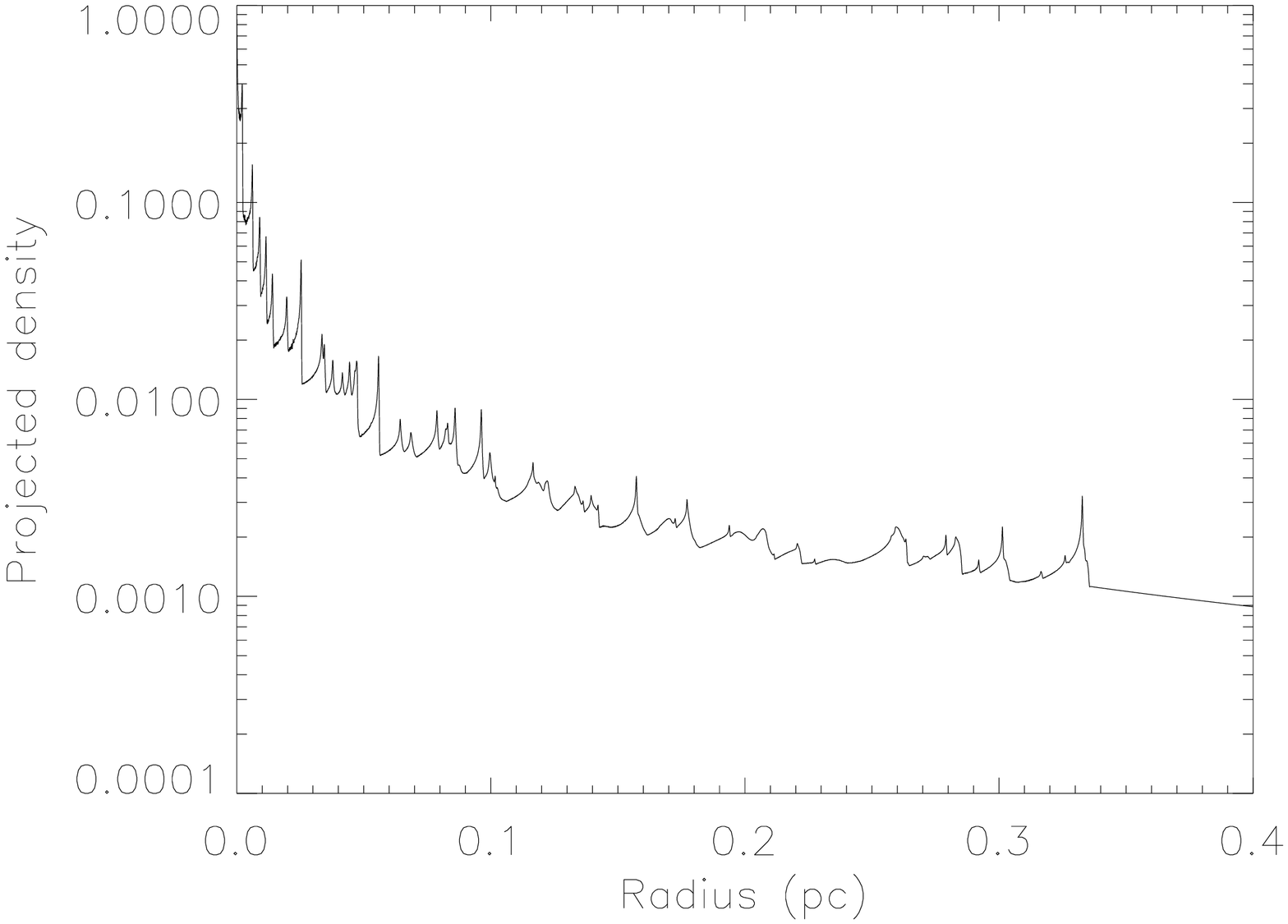}
\caption{Density profile as a function of radius (top) and projected on the 
sky (bottom)}
\label{InitDens}
\end{figure}

\section{Simulations}

We performed five simulations for different initial density distributions
and stellar masses. In this section we describe the results of one and
refer to the results of the other simulations only when they help us
understand certain features of behaviour.

\subsection{The initial density distribution}

In Sect.~2.1 the mechanism of the mass loss variations during the AGB was
discussed. From the two-fluid dust-gas simulations of \citet{Simisetal2001}
we took one result as the input for our simulation. These mass loss and
velocities variations were calculated using the properties of IRC+10216
\citep{Wintersetal94}: a stellar mass of 0.7~\msun, an effective temperature
of $1.98\times 10^3$~K, luminosity $2.4 \times 10^4$~$L_\odot$, and a
carbon/oxygen ratio of 1.4.

Our main aim is to follow the post-AGB evolution of the {\it type\/} of
variations in density and velocity which are being produced by the two-fluid
calculations. We therefore scale the outcome of the two-fluid simulation to
obtain a more typical AGB wind with an average mass loss rate of $2.5\cdot
10^{-5}\ M_\odot$~yr$^{-1}$, and an average velocity of 10~km~s$^{-1}$. This
means that the AGB and post-AGB simulations are no longer physically
connected and the parameters used for the AGB star listed above are no longer
relevant for the results.  What we are effectively doing is using the
two-fluid calculations to generate a spectrum of density and velocity
variations which at least has some physical basis (rather than assuming
sinusoidal variations), and apply this to our AGB wind. This is also
motivated by the fact that the quantitative outcome of the two-fluid
simulation is quantitatively uncertain since the model does not include
effects such as pulsation, evolution of the AGB star, etc.

The computational grid of the dust-gas simulations extended to a radius of
only $1.8\cdot 10^{16}$~cm (1200 AU) and to be able to follow the long term
(post-AGB) evolution we first had to fill a larger space with mass loss and
velocitiy fluctuations. For this we extracted the density and velocity as a
function of time at a radius of $5.0\cdot 10^{15}$~cm (333 AU), and used this
as an inflow condition for a period of 31000 years, by which time the
fluctuations fill most of the grid. The temperature at the inner
boundary is taken from the two-fluid calculation, and we extrapolate outward 
using a power law, $T \propto r^{-2(\gamma-1)}$, with an imposed minimum 
value of 10~K. The actual choice for the temperature profile is not very 
important for the dynamics, since the gas is flowing out supersonically. 

As the fluctuations move out, the distances between minima in density become
larger. This is due to the merging of shells, since these do not have the
same velocity. Figure~\ref{InitDens} shows the structure of the AGB wind
31000 years after we introduced the variable wind. One sees that the spacing
of the rings increases with distance from the star. When this density pattern
is projected on the sky the highest frequency structures disappear. As a
consequence it is easier to distinguish the shells.

\begin{figure*}
\centerline{\includegraphics[angle=90,height=105mm,clip=]{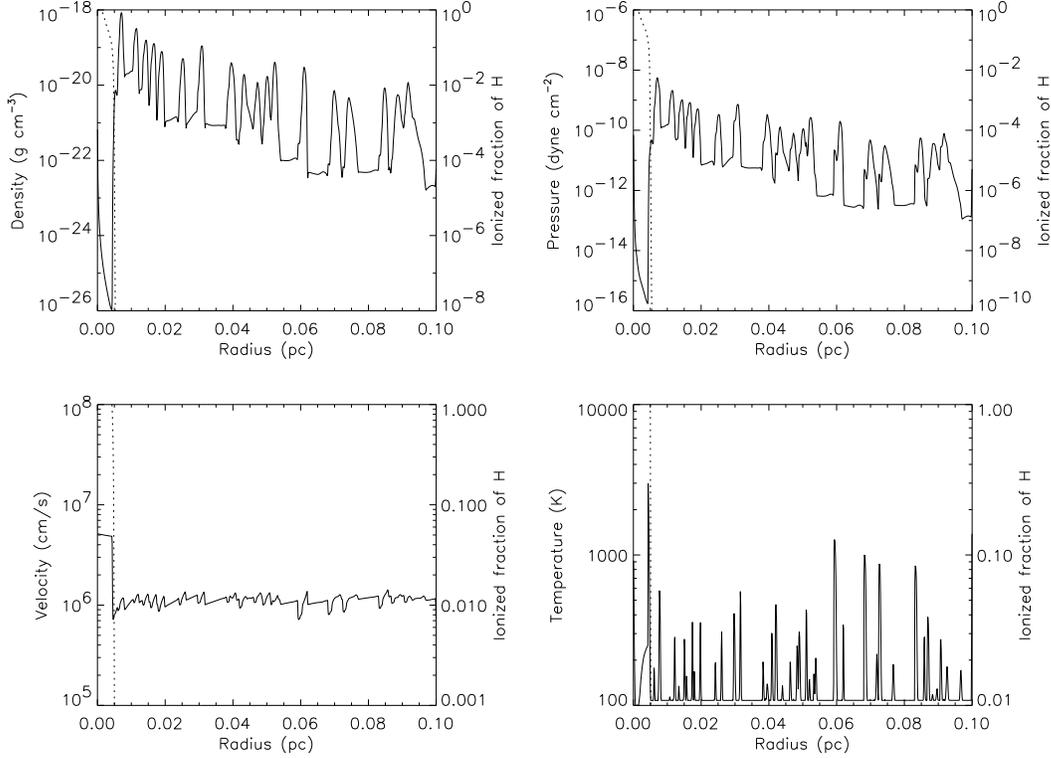}}
\caption{Result after 400 years: Density (top left), pressure (top right),
velocity (bottom left) and temperature (bottom right). The dotted line 
shows the ionized fraction of hydrogen.}
\label{Denst400}
\end{figure*}

\subsection{The dynamical evolution of the rings}

Our standard model is for a star with a Zero Age Main Sequence (ZAMS) mass of
1.0~$M_\odot$ and post-AGB mass of 0.565 $M_\odot$. We call $t=0$ the start
of this simulation, which corresponds to the start of the post-AGB evolution.
The simulation follows the evolution of the circumstellar material over a
time span of 15000 years and is performed on a grid of 6000 points. The
post-AGB wind first collides with the slow AGB wind at a distance of 
$5\times 10^{15}$~cm, the inner edge of the grid lies at $10^{14}$~cm
(6.6 AU).

In Fig.~\ref{Denst400} the result after 400 years is shown. The effective
temperature of the star is about 6000 K. There are enough energetic photons
to ionize the fast wind. Because the velocity of the fast wind is 50 km/s and
the mass loss rate of the star $1.6\cdot 10^{-12}\ M_\odot$~yr$^{-1}$, the
temperature in the shock, moving into the fast wind (the inner shock), is not
that high, $\sim$~3000 K. The pressure in the inner shock is actually lower
than in the slow wind. As a consequence, the slow wind dissipates slightly
inwards. The ionization front lies at the interaction between the fast and
the slow wind.

\begin{figure*}
\centerline{\includegraphics[angle=90,height=105mm,clip=]{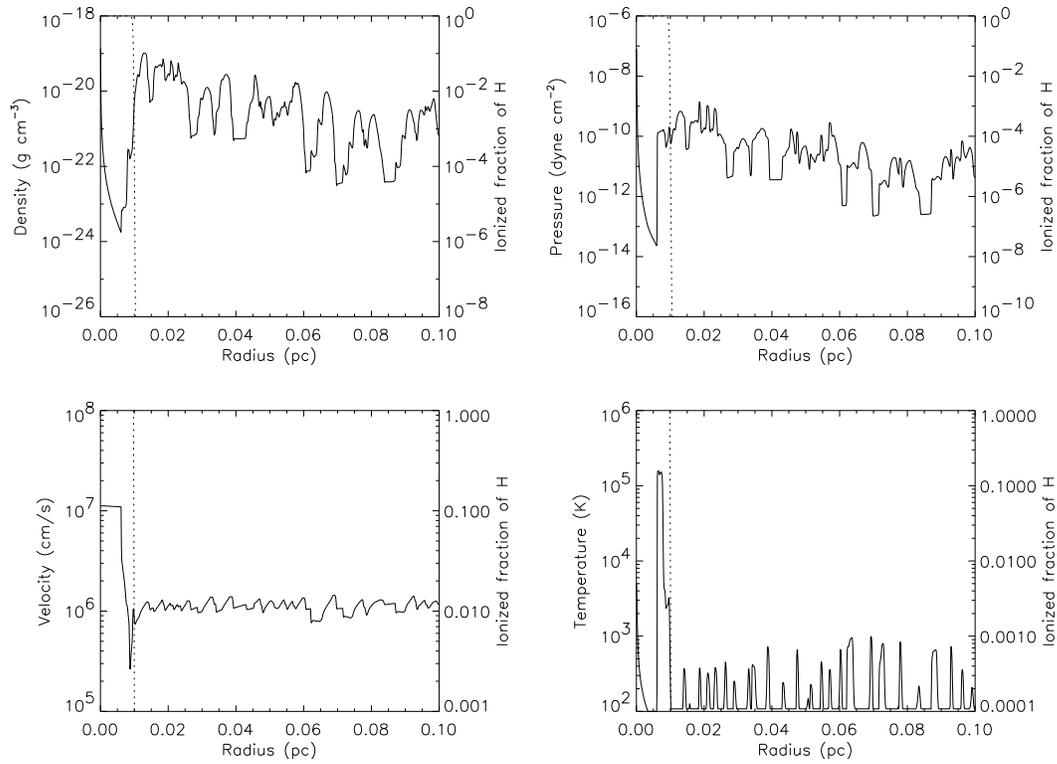}}
\caption{Result after 900 years: Density (top left), pressure (top right),
velocity (bottom left) and temperature (bottom right). The dotted line 
shows the ionized fraction of hydrogen.}
\label{Denst900}
\end{figure*}

Sixhunderd years later (Fig. \ref{Denst900}), the effective temperature is
8200 K, and the mass loss rate went up by three orders of magnitude to $\sim
10^{-9} M_\odot$ ~yr$^{-1}$, with a velocity to somewhat over 100 km/s. As a
consequence, the temperature in the hot bubble is well over $10^5$~K. The
pressure is now the same as in the slow wind. One can also see the ionization
front moving into the slow wind. The temperature here is about 2500 K. The
gas in front of the ionization front is not shocked yet and has a temperature
around 100 K. One also sees that the shells merge while they move outward. At
the place where the shells interact the temperature rises to a few hunderd
degrees.

\begin{figure*}
\centerline{\includegraphics[angle=90,height=105mm,clip=]{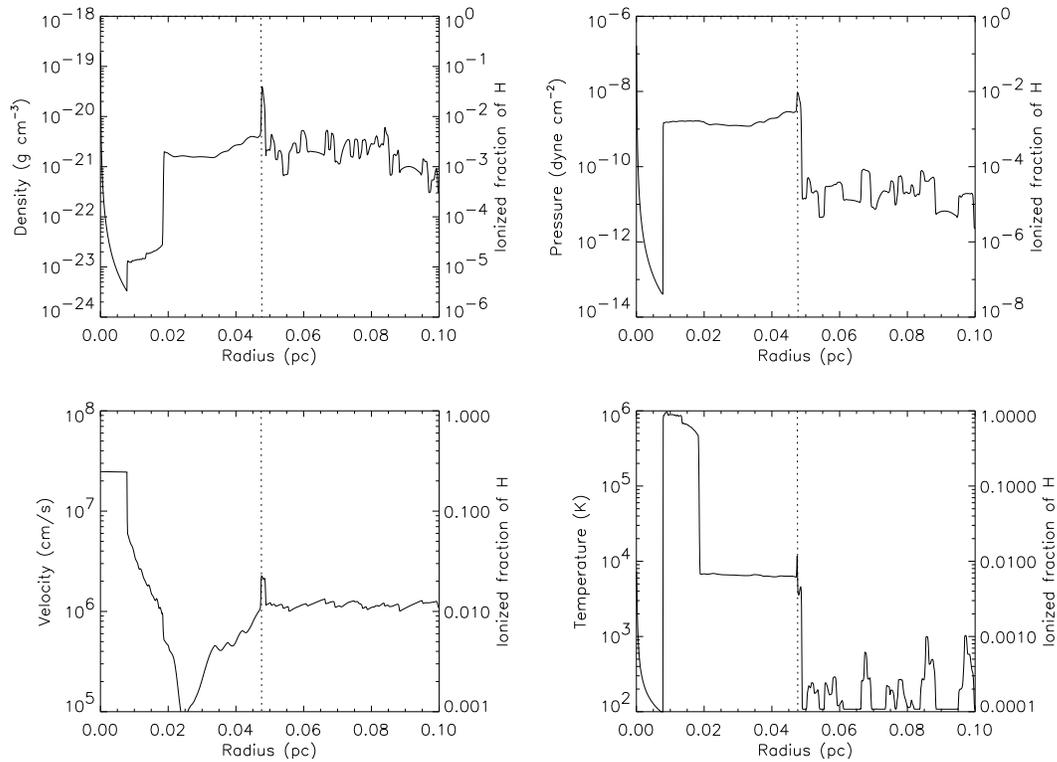}}
\caption{Result after 3000 years: Density (top left), pressure (top right),
velocity (bottom left) and temperature (bottom right). The dotted line 
shows the ionized fraction of hydrogen.}
\label{Denst3000}
\end{figure*}

The ionization front has moved further out to 0.15 pc, 2100 years later (see
Fig.~\ref{Denst3000}), when the star's effective temperature has risen to
20000 K. The part of the wind that is ionized has almost no density
variations. Since the pressure of the gas in the ionized region is higher,
the gas in front of the ionization front is pushed outward. This gas piles up
in a thin shocked shell and with a temperature somewhat lower than the
ionized region. This is typical of a D-type ionization front, where the speed
of the front is slow enough for the gas to react to the increased pressure of
the ionized region. This configuration also produces the typical positive
velocity gradient in the ionized material, found in many PNe.

\begin{figure*}
\centerline{\includegraphics[angle=90,height=105mm,clip=]{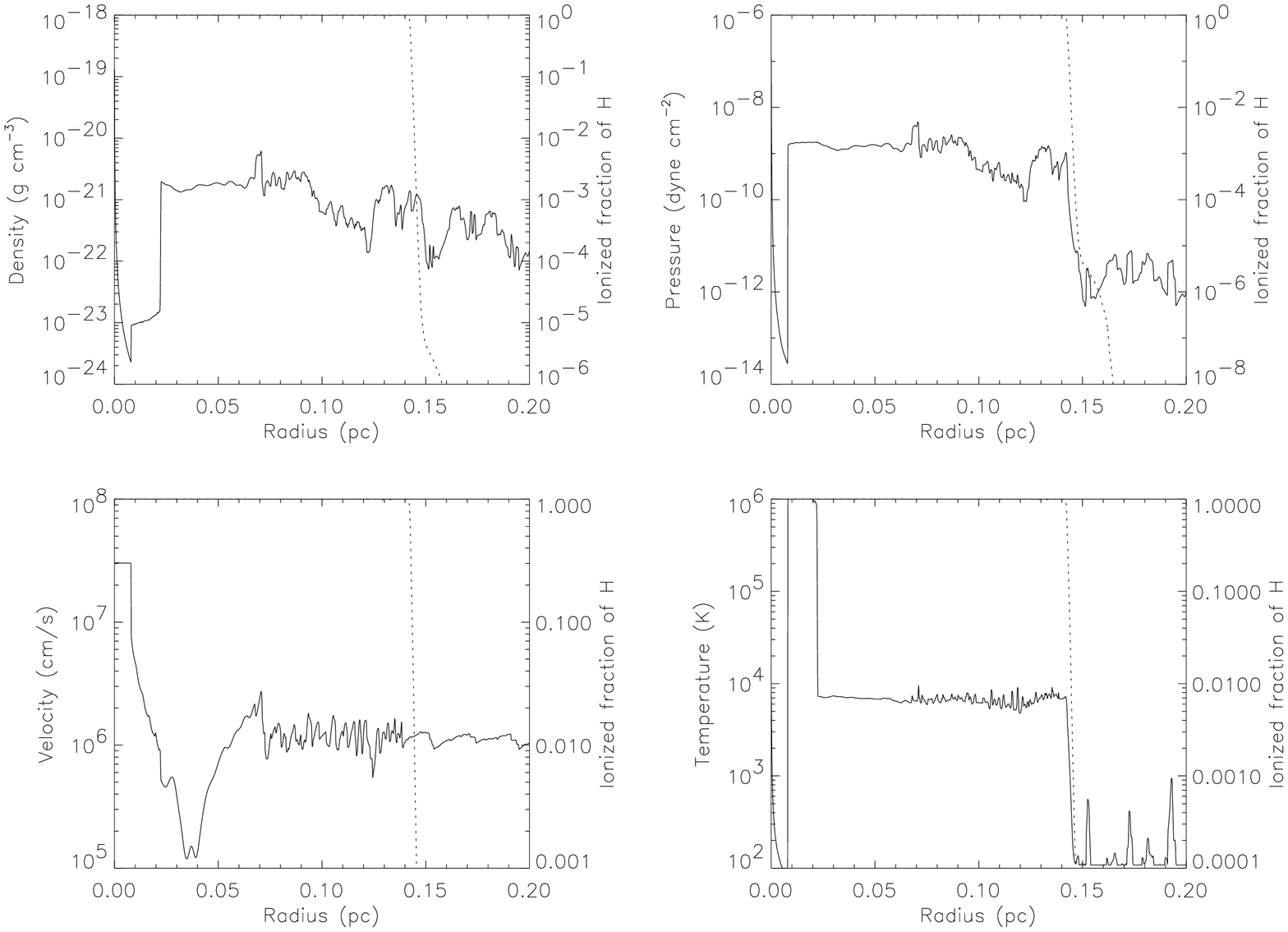}}
\caption{Result after 3700 years: Density (top left), pressure (top right),
velocity (bottom left) and temperature (bottom right). The dotted line 
shows the ionized fraction of hydrogen.}
\label{Denst3700}
\end{figure*}

Almost 3700 years after the end of the AGB (see Fig. \ref{Denst3700}) the
star's effective temperature ($\sim$~24000 K) reaches the point that the
ionization front breaks through. In a few hundred years the whole nebula gets
ionized and goes from ionization bounded to density bounded. The temperature
in the ionized part of the slow wind rises to about 7000 K, and as a
consequence the pressure also rises.  The pressure being proportional to
$\rho T$, follows the density profile and large pressure differences of an
order of a magnitude occur. The gas tries to equalize this out, the effect of
which can be seen in the velocity field. In the ionized region the velocity
variations are over a shorter distance and have a larger amplitude than in
the neutral region.

\begin{figure*}
\centerline{\includegraphics[angle=90,height=107mm,clip=]{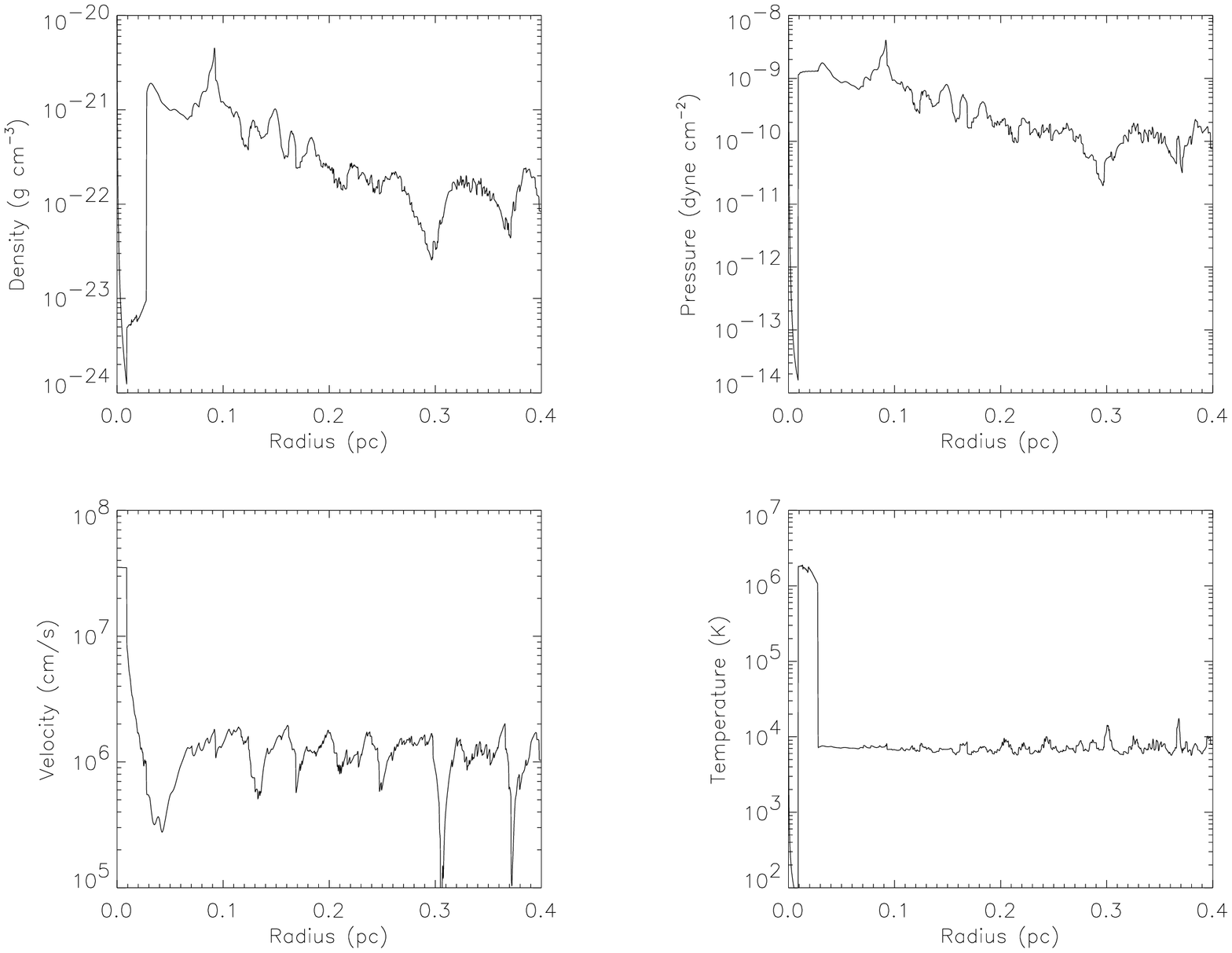}}
\caption{Result after 4500 years: Density (top left), pressure (top right),
velocity (bottom left) and temperature (bottom right). All the hydrogen is
ionized.}
\label{Denst4500}
\end{figure*}

At 4500 years (see Fig. \ref{Denst4500}) the velocity variations have
increased, and the velocity in the rings now range from 1 to 20~km~s$^{-1}$.
These variations are much larger than in the period before ionization. To
check the dependence on the initial velocity fluctuations, we did a
simulation with the same initial density variations but with a constant
velocity field. We found essentially the same velocity variations developing
after ionization. This shows that the initial velocity variations are
unimportant in this phase.

\begin{figure*}
\centerline{\includegraphics[angle=90,height=107mm,clip=]{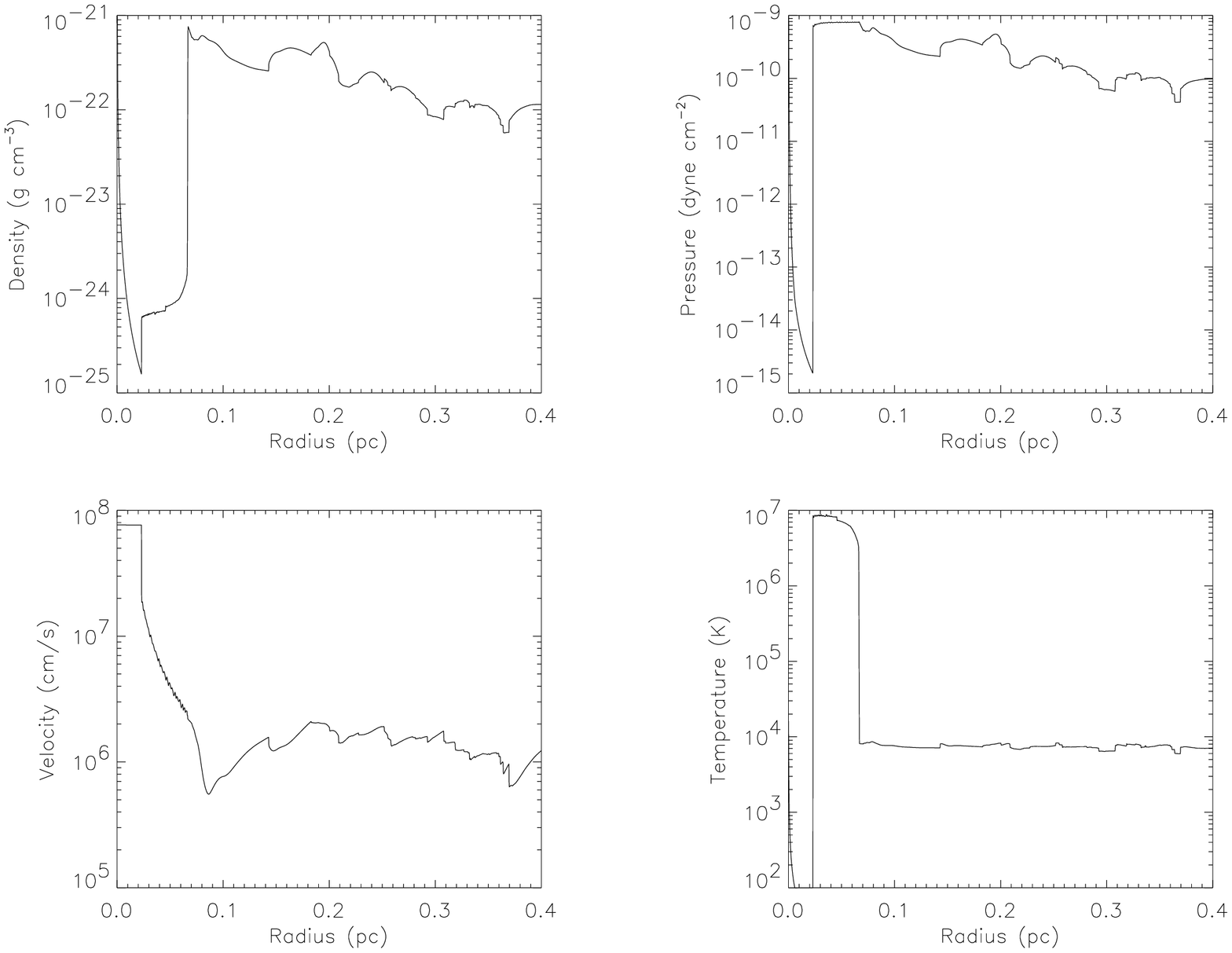}}
\caption{Result after 8000 years: Density (top left), pressure (top right),
velocity (bottom left) and temperature (bottom right).}
\label{Denst8000}
\end{figure*}

Considering the situation 3500 years later (see Fig.~\ref{Denst8000}), we
find that the density and velocity variations are still visible. These
variations eventually equalize out, but even after 15000 years some 
traces of it can still be distinguished.

\begin{figure*}
\centerline{\includegraphics[angle=90,height=105mm,clip=]{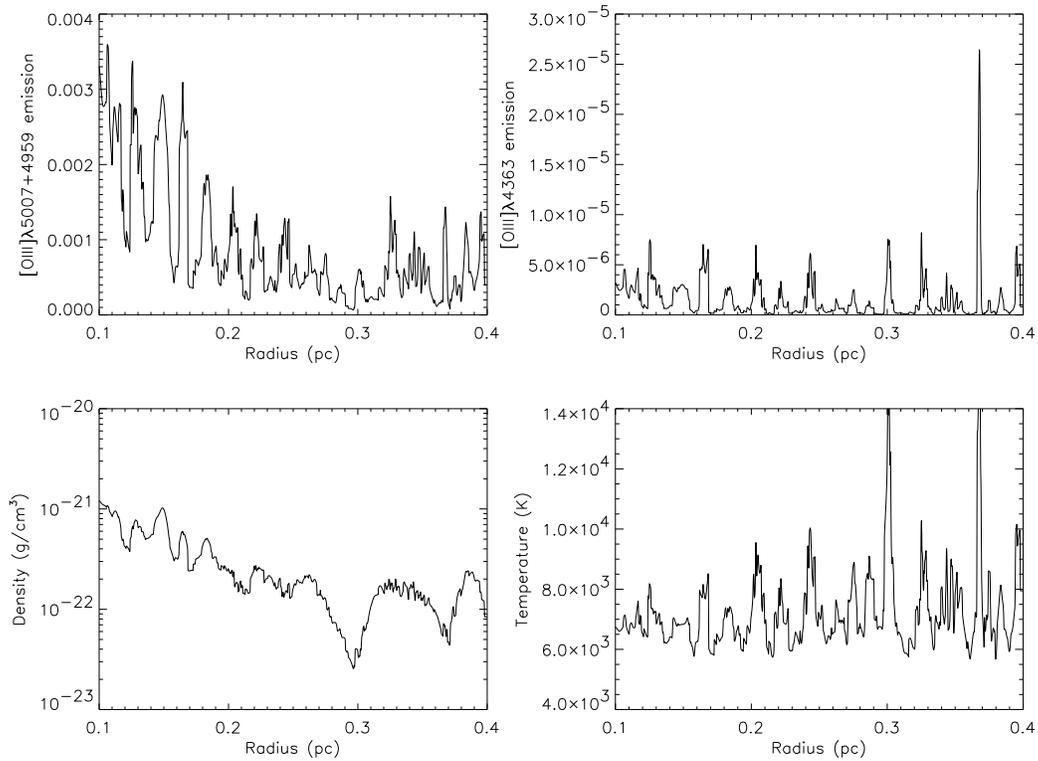}}
\caption{Emission from [OIII]$\lambda$4959 + $\lambda$5007 (top left),
$\lambda$4363 (top right) in arbitrary units, density profile (bottom left) 
and temperature profile (bottom right) at $t$=4500 years}
\label{oiii4500}
\end{figure*}

\section{Synthesized observations}

\subsection{Emission lines and temperature}

A radial density profile does not necessarily correspond to what would be
observed. In order to compare to the observations, we calculated emissivities
of certain lines, in [O III]~5007\AA, the line in which the rings in
NGC 6543 are most easily observed, as well as the [O III]4363\AA\ line, whose
ratio with the 5007 line gives us an `observed temperature'. We calculated
the line emissivity from the density, temperature, and O$^{2+}$ concentration
at every point.  The emission is calculated conform Sect.~3.5 and Tables
3.4 and 3.8 of \citet{Osterbrock1989}.

The radial [O III]5007\AA\ and 4363\AA\ emissivities at $t$=4500 years are
shown in Fig.~\ref{oiii4500}, together with the density and temperature. One
can see that the density profile does not correspond to the [O~III]
profiles. This is due to the temperature dependence of the [O III] line which
gives more weight to points of higher temperature, especially for the
4363\AA\ line.

\begin{figure*}
\centerline{\includegraphics[angle=90,height=52mm,clip=]{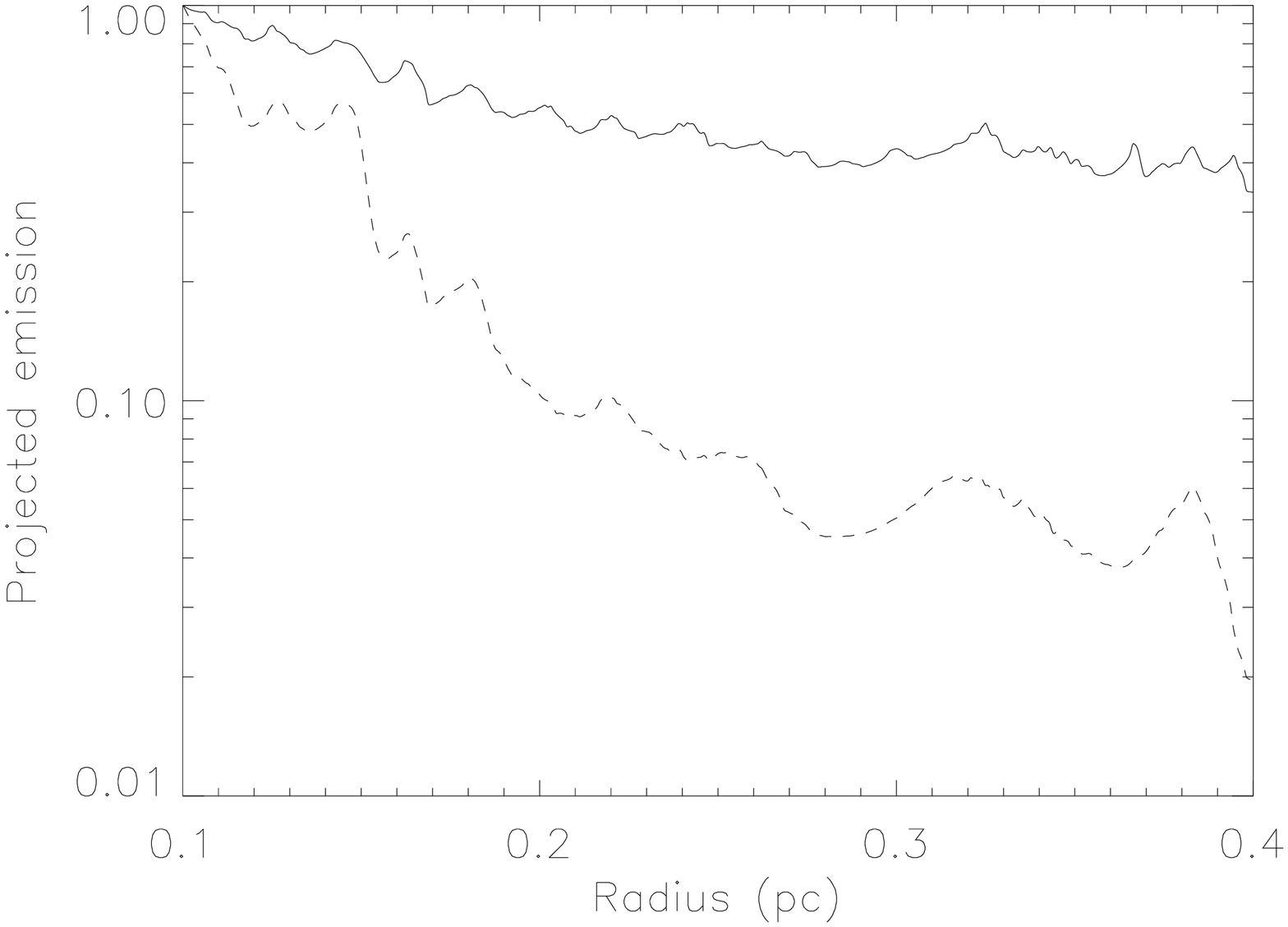}
\includegraphics[angle=90,height=52mm,clip=]{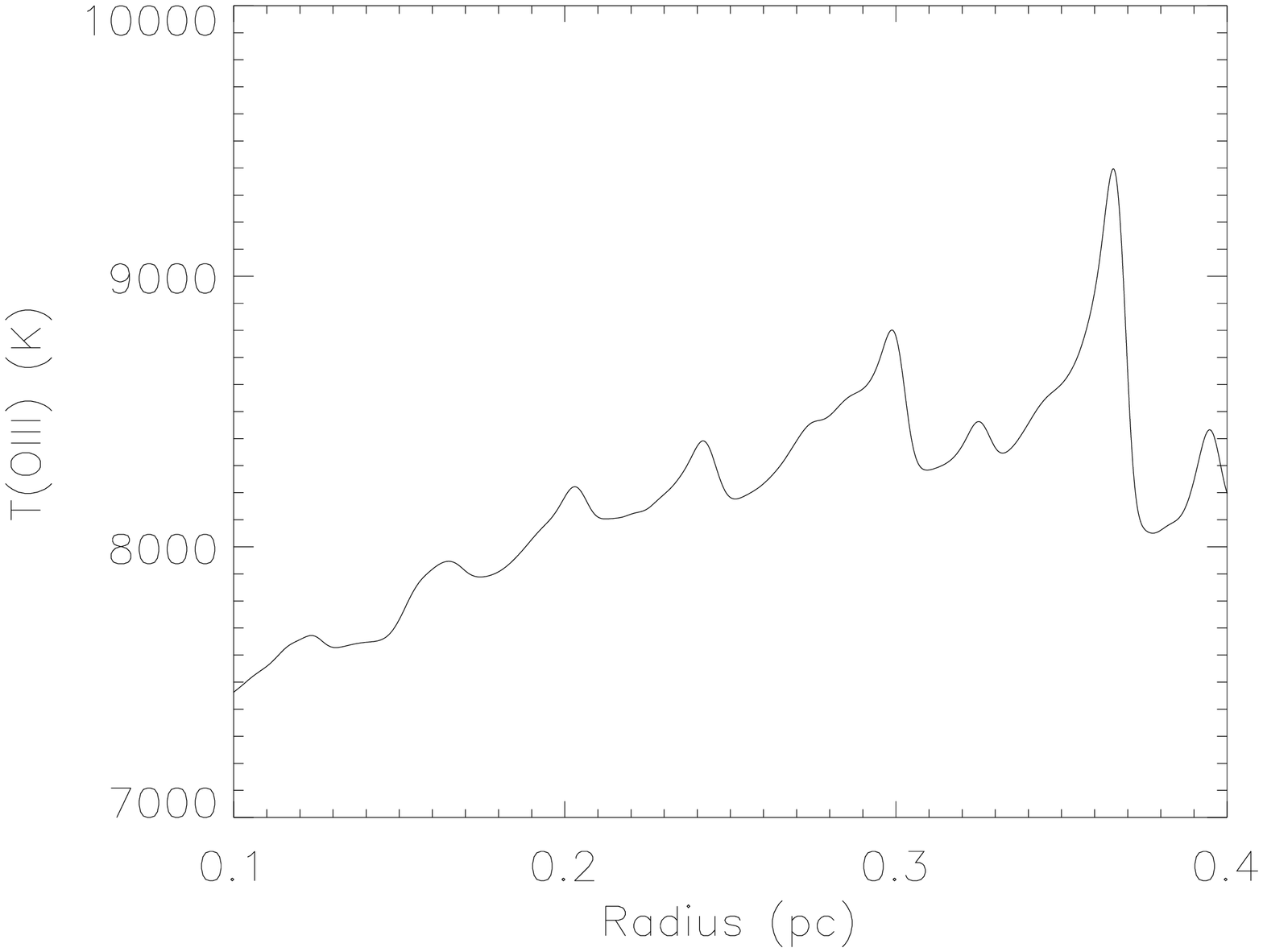}}
\caption{Left: The projected emission [OIII] (solid) and H$\alpha$ (dashed) 
in arbitrary units. Right: Projected temperature. $t$=4500 years.}
\label{ProjEm4500}
\end{figure*}

To compare to observations, we projected the emissivities onto the sky, and
convolved them with a gaussian of FWHM corresponding to 0.1 arcsec at the
distance of NGC 6543. We then took the ratio between the 5007 and 4363\AA\
lines to determine the observed temperature. These projected emission and
temperature profiles for the rings area are shown in Fig.~\ref{ProjEm4500}. 

The [O III] emission profile shows clear maxima and minima. The spacing
between the maxima is about 0.02 pc. The dashed line shows the same in
H$\alpha$. Clearly the two have different radial dependencies. Figure~2
of \citet{Balicketal2001} shows how the rings are much more pronounced in the
[O~III] profile than in the H$\alpha$ profile and how the H$\alpha$ emission
falls off more rapidly, something which our model at least qualitatively
reproduces.

The observed temperature varies between 7500~K and 10000~K, but most of this
is due to a small outward gradient. Since the radial temperature profile does
not show a temperature gradient, we must conclude that this is a projection
effect. This is an interesting side-result, showing another way in which
deriving electron temperatures from line ratios can be somewhat
misleading. The true local variations are mostly of order 500~K. The biggest
variations in temperature are seen in the outer part of the nebula, where
most shocks occur at this time.

\begin{figure*}
\centerline{\includegraphics[angle=90,height=52mm,clip=]{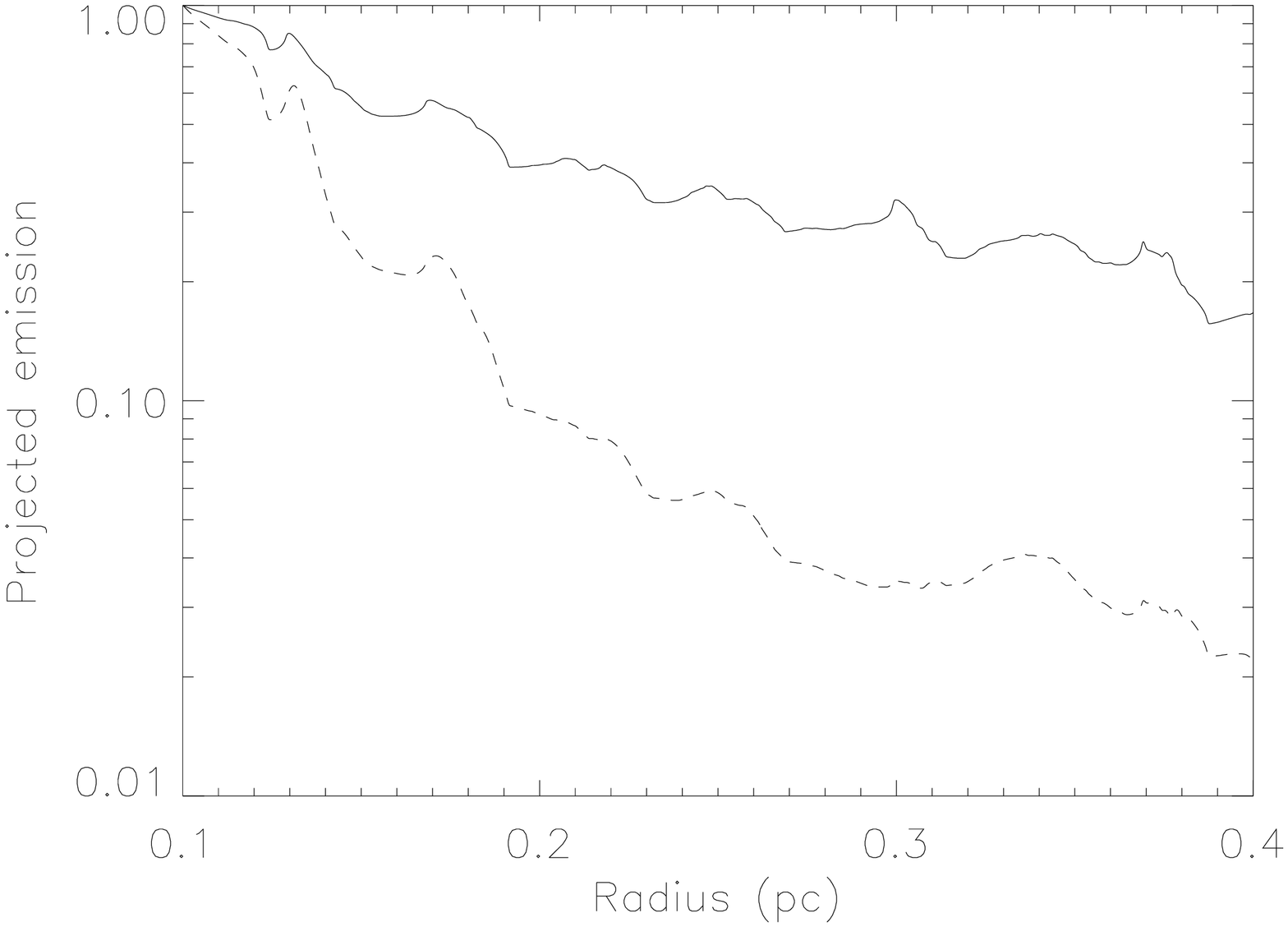}
\includegraphics[angle=90,height=52mm,clip=]{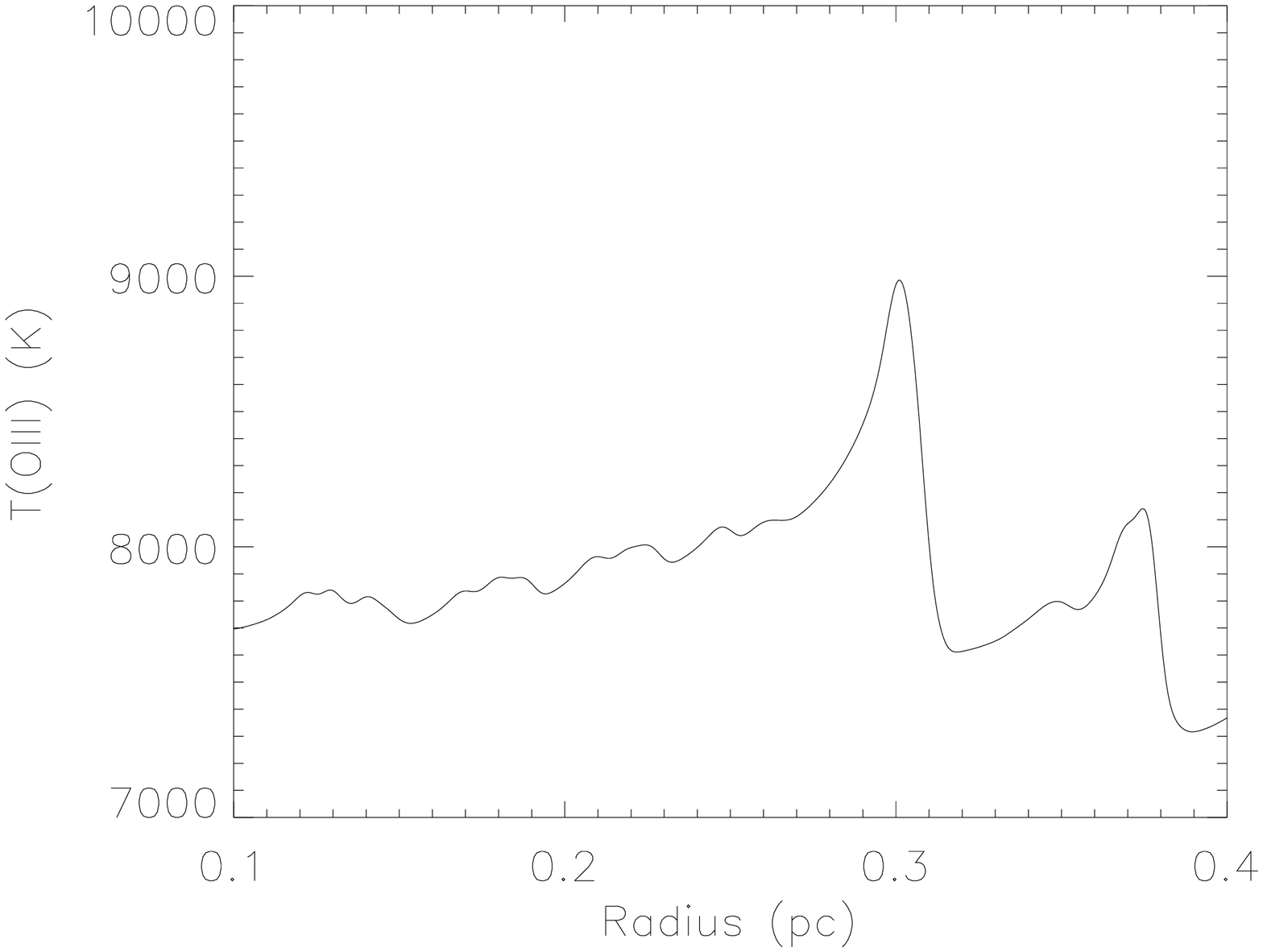}}
\caption{Left: The projected emission of [OIII] (solid) and H$\alpha$ (dashed) 
in arbitrary units. Right: Projected temperature. $t$=5500 years.}
\label{ProjEm5500}
\end{figure*}

One thousand years later (Fig.~\ref{ProjEm5500}), the temperature variations
have decreased substantially. The largest variations are still seen in the
outer region, where the shells are merging. Due to this merging the spacings
between maxima have become larger.

\begin{figure*}
\centerline{\includegraphics[angle=90,height=52mm,clip=]{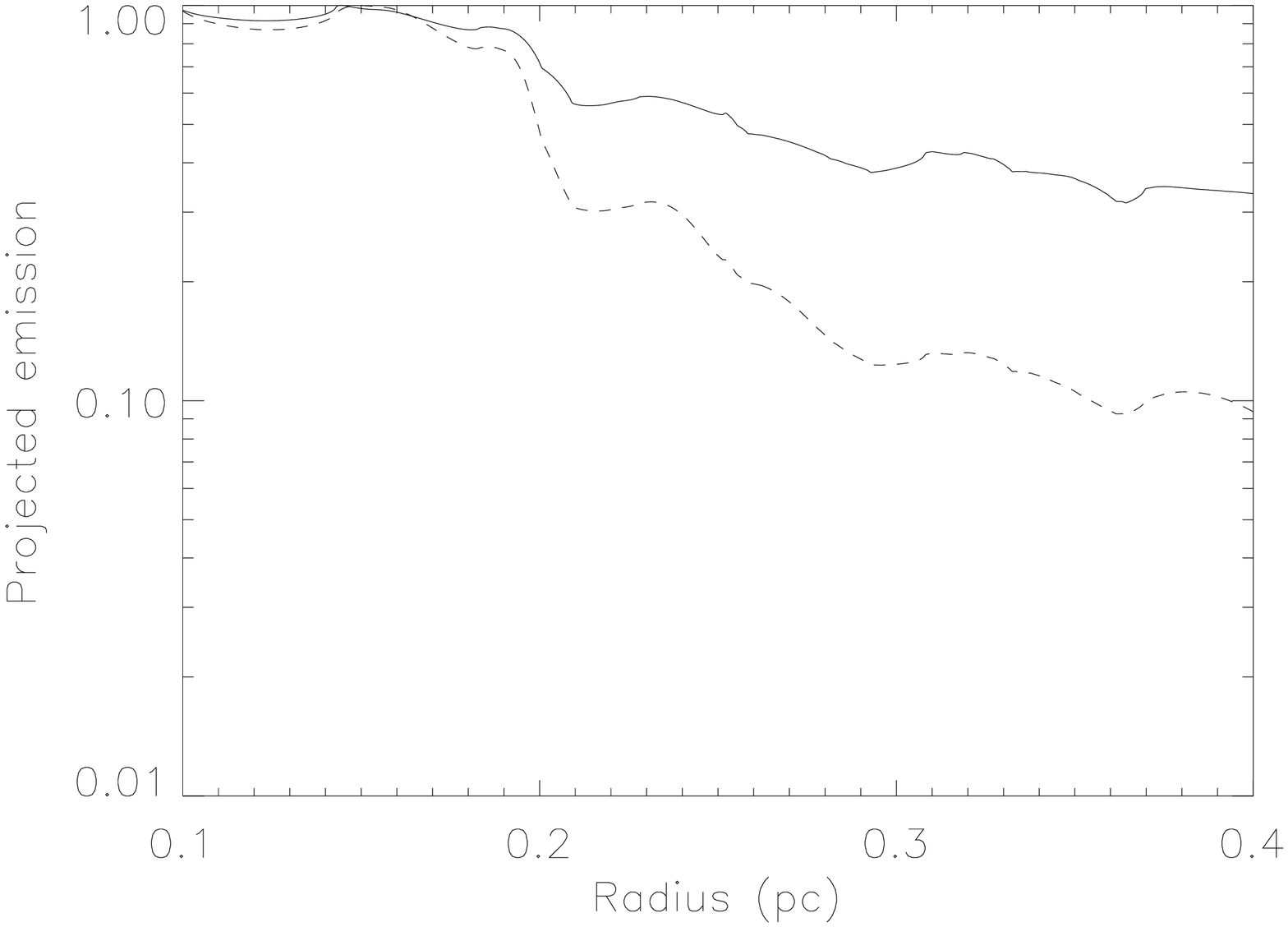}
\includegraphics[angle=90,height=52mm,clip=]{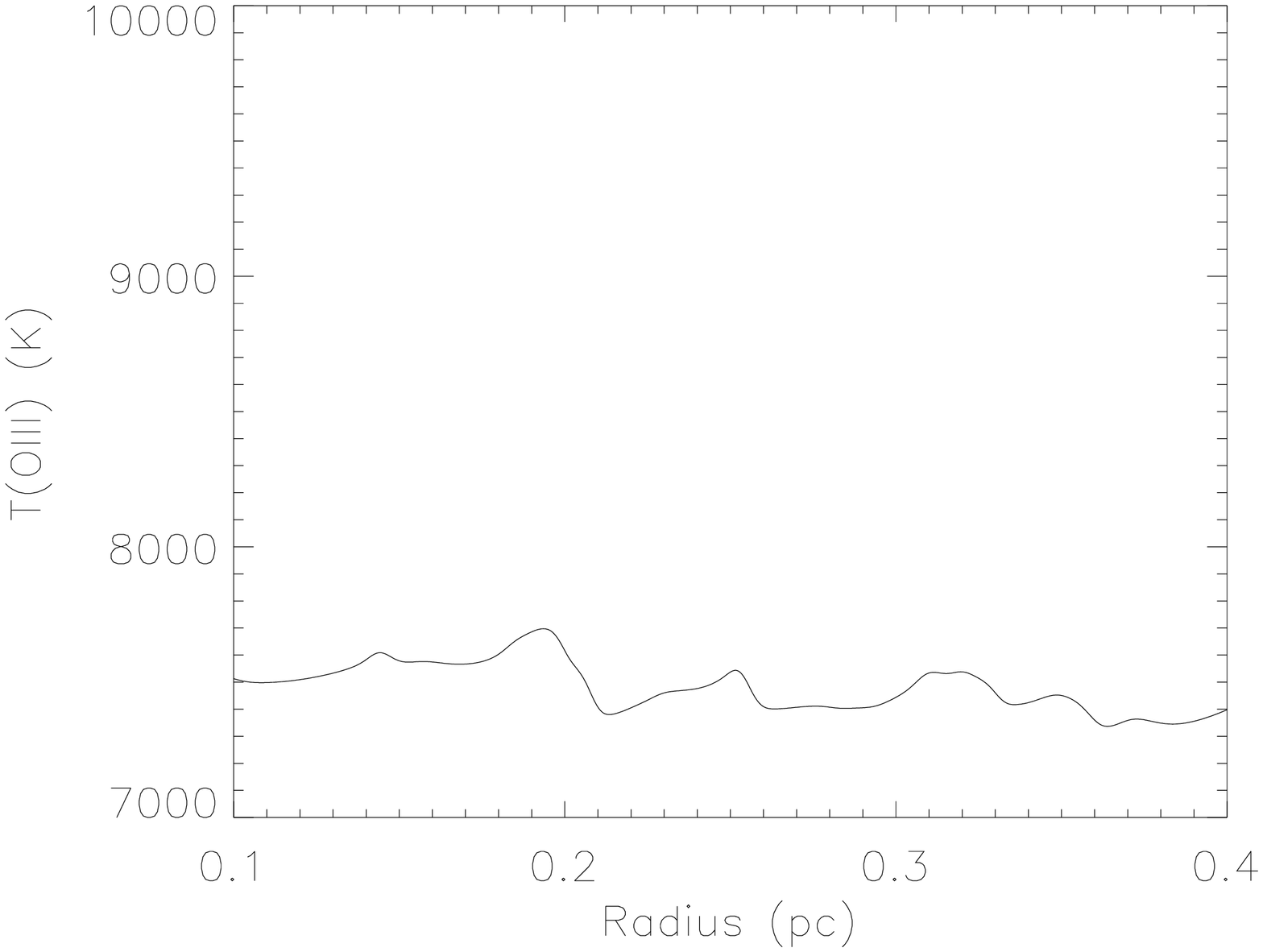}}
\caption{Left: The projected emission of [OIII] (solid) and H$\alpha$ (dashed)
 in arbitrary units. Right: Projected temperature. $t$=8000 years.}
\label{ProjEm8000}
\end{figure*}

At $t$=8000 years almost no temperature variations are seen, the amplitude
has dropped below a few hunderd Kelvin (Fig.~\ref{ProjEm8000}). Also in the
density profile the rings have become less noticeable, although they can
still be distinguished.

\subsection{Lineprofiles}

\begin{figure*}
\centerline{\includegraphics[angle=90,height=105mm,clip=]{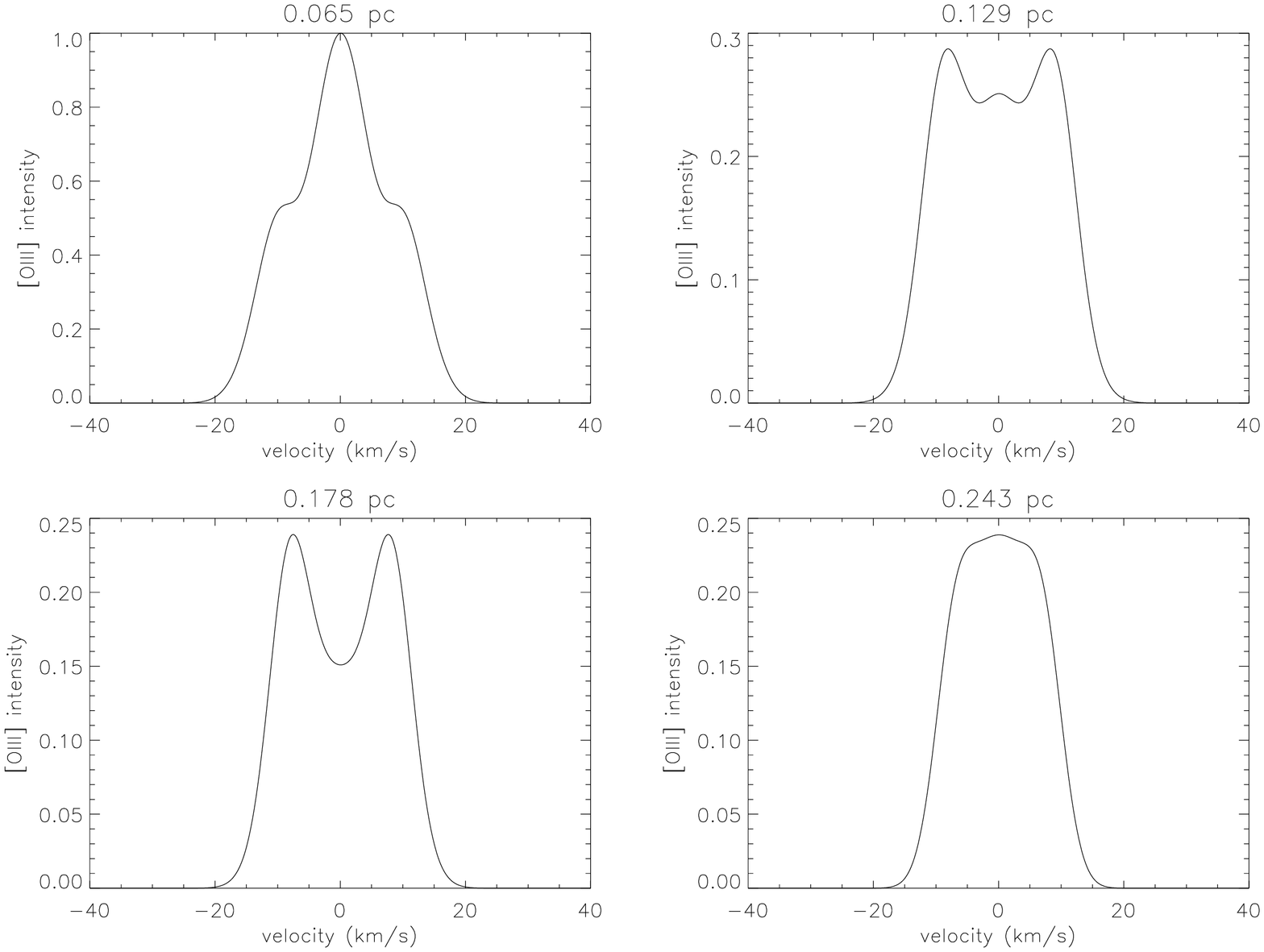}}
\caption{Lineprofiles for [OIII] at several distances from the star. 
The lines are convolved with a gaussian of 6 km~s$^{-1}$ FWHM.}
\label{plott4500}
\end{figure*}

Combining the emissivities with the velocity information from the simulation,
allows us to derive line shapes. We present line shapes at different
positions along the rings in Fig. \ref{plott4500}. Since the temperature of 
the gas is around 7500 K, the lines are also thermally broadened which we
simulated by convolving with a gaussian of 6 km~s$^{-1}$ FWHM. The effects of
seeing or slit width were not taken into account here.

One sees how the profiles are quite broad, up to 30 km~s$^{-1}$ FWHM. This is
due to the fact that at a given position one observes several overlapping
shells producing different velocity components. The profiles are clearly not 
gaussian, so the effect that the emission comes from different shells is 
visible.

These line profiles match the reported observations quite
well. \citet{Bryceetal92} and \citet{Balicketal2001} describe (but do not
show) wide ($\sim 30$~km~s~$^{-1}$) line profiles in the region of the rings.
The widths match our model line profiles well, the low S/N of the
spectroscopy of the faint rings makes further comparison difficult.

\section{Discussion}

The comparison with the observations of NGC~6543 in the previous section
shows that our model matches a number of key observations:
\begin{enumerate}
\item The rings are clearly visible in the [O~III] images.
\item The H$\alpha$ and [O~III] emissivities have slightly different radial
  profiles.
\item The line profiles are broadened to $\sim30$~km~s~$^{-1}$.
\end{enumerate}
This shows that our model gives a good description of the rings.

Comparing the spacing from our models with what is found in NGC 6543, we find
a small descrepancy. Our spacing is 0.02~pc, about twice as large as in the
observations (0.01--0.015~pc), assuming the distance to NGC 6543 is reliable.
However, the exact spacing of the rings is influenced by many factors. First,
the original spacing in the variable AGB mass loss depends on the density
scale height at the base of the circumstellar envelope. Second, over time
shells merge, increasing the spacing, and hence making it a function of
time. A faster evolution will thus produce shorter distances between the
shells.

The derived observable temperature (from the [O III]4363/5007 ratio) is found
to vary somewhat, but not as much as claimed by \citet{Hyungetal2001},
supporting the results of \citet{Balicketal2001} for an electron temperature
comparable to those in the core nebula. We predict temperature variations of
not much more than 1000~K, and perhaps a positive outward temperature
gradient, depending somewhat on the phase NGC~6543 is in.

Our results show that the rings are a transient phenomenon, persisting not
much longer than $\sim$3000~years after photo-ionization. This seems to be in
line with the observed fact that most rings are found in their neutral state 
around proto-PNe
or young PNe (e.g.~NGC 7027). In this view the ionization in NGC 6543 is rather
recent. In Hb~5 it is not known whether the rings are seen in scattered or
intrinsic emission.

Another prediction from our models is that one will not find rings in
so-called `attached haloes' or shells around PNe, since these are thought to
have been produced by the action of a D-type ionization front
\citep{MartenSchonberner, RHPNIII}, which as we saw, erases the density
variations quite efficiently.

\subsection{Origin of the rings}
To produce our models we started with density and velocity variations produced
through the mechanism suggested by \citet{Simisetal2001}, which we slightly
modified by taking a lower velocity, and scaling the density to obtain a
certain average mass loss rate. Our simulations showed that for the structure
during the PN phase, the initial velocity fluctuations are irrelevant, and so
the important property is the nature of the density fluctuations. Studying the
results of a number of simulations, we found that the key property is the
amplitude of the density variations. At the start of the post-AGB phase, these
have to be roughly one order of magnitude. Only then ionization leads to large
enough pressure fluctuations to result in what could be loosely described 
as `supersonic
turbulence', which helps the longevity of the rings. Without the resulting
shock waves the rings diffuse away in a few hundred years after ionization.

We would therefore expect that using density variations with similar
amplitudes would give more or less the same results. In the model of
\citet{Guille2001}, for reasonable values of the magnetic field, the density
initially varies less than in our simulations. However, since these shells are
`frozen in', they do not merge, and conserve their density contrast. At
the time when the ionization front breaks through, the density contrast is
likely to be similar to the one in our simulation. We therefore expect the
rings from the MHD model to evolve similarly during the PN phase, and survive
for about a few thousand years after ionization, producing broad line profiles
as the flow is slowly erasing the rings.

Note that the magnetic pressure is able to support the rings only during the
AGB and proto-PN phases. Upon ionization, the magnetic pressure becomes
negligible in comparison to the gas pressure, and there is no real difference
between the two models. This rather limits the usefulness of the magnetic
field, since it is not clear whether it is really needed to maintain the
fluctuations up to the time of ionization.

Binary interaction as proposed by \citet{MastroMorris99} is probably not an
explanation for the rings seen in planetary nebulae. The density contrasts they
find in their models are much too low to survive for a long time after being
ionized, and in fact already seem to be diffusing within a couple of
orbital periods.

\section{Conclusions}  

We numerically followed the evolution of AGB mass loss fluctuations during 
the post-AGB phase, paying special attention to the effects ionization has.
We find that using mass loss fluctuations as produced by the mechanism
described by \citet{Simisetal2001} easily survive up to the time ionization
sets in for stars of ZAMS mass of 1~\msun or larger. Upon ionization, the
pressure fluctuations lead to shocks, which raises the average absolute
velocity of the rings, and locally raises the temperature. After a couple of
thousand years the rings will have mostly disappeared.

The ionized rings have the following observational characteristics:
\begin{itemize}
\item The rings are more pronounced and can be seen further out in [O III]
  then in H$\alpha$.
\item Only slightly higher electron temperature ($\pm 1000$~K) than the rest
  of the PN, as measured by the [O III] line ratios.  
\item Broad line profiles due to the turbulent velocity field and the many
  overlapping shells.
\end{itemize}

We therefore predict that the electron temperature from accurate line ratios
of ionized rings will show only a small difference with the electron
temperature of the core nebula, showing the result from \citet{Hyungetal2001}
to be spurious. We also predict that high signal-to-noise spectroscopy may
reveal non-gaussian line shapes in the ring region. Lastly, according to our
description, rings will only be found in either neutral or recently ionized
haloes.

We found this type of evolution to be quite general, provided that the density
fluctuations at the time of ionization are larger than a factor of 10. For
lower values the rings diffuse away through gentle sound waves.

Using this requirement the wide binary model as published by
\citet{MastroMorris99} disqualifies as an explanation for the rings in ionized
PNe. The MHD model of \citet{Guille2001} and the dust-driven wind instability
from \citet{Simisetal2001} both fulfill the requirement. However, there is 
some dispute on the choice of period for the magnetic cycle of AGB stars,
making the dust-driven wind instability model (with no imposed period), the
more likely explanation.

\begin{acknowledgements} 
The research of GM has been made possible by a fellowship of the Royal 
Netherlands Academy of Arts and Sciences. 
\end{acknowledgements} 

\bibliographystyle{aa} 
\bibliography{3404.bib} 

\end{document}